\documentstyle[aps,prb,multicol]{revtex}
\input{psfig}
\begin{document}
\title{Shot noise suppression in multimode ballistic Fermi conductors}
\author{Y. Naveh, A. N. Korotkov, and K. K. Likharev}
\address{Department of Physics and Astronomy \\
State University of New York, Stony Brook, NY 11794-3800}
\date{December 11, 1998}
\maketitle

\begin{abstract}
We have derived a general formula describing current noise in multimode
ballistic channels connecting source and drain electrodes with Fermi
electron gas. In particular (at $eV\gg k_{B}T$), the expression describes
the nonequilibrium ''shot'' noise, which may be suppressed by both
Fermi correlations and 
space charge screening. The general formula has been applied to an
approximate model of a 2D nanoscale, ballistic MOSFET. At large negative
gate voltages, when the density of electrons in the channel is small, shot
noise spectral density $S_{I}(0)$ approaches the Schottky value $2eI$, where 
$I$ is the average current. However, at positive gate voltages, when the
maximum potential energy in the channel is below the Fermi level of the
electron source, the noise can be at least an order of magnitude
smaller than the 
Schottky value, mostly due to Fermi effects.
\end{abstract}

\renewcommand{\tt}{{\bf t}} \renewcommand{\Re}{{\rm Re}} \renewcommand{\[}{
\begin{equation}} \renewcommand{\]}{\end{equation}} 

\begin{multicols}{2}

Nonequilibrium current fluctuations (''shot noise'') in ballistic
conductors have been studied extensively in recent years both
theoretically\cite{Khlus 87,Lesovik 89,Buttiker 90,Landauer 91} and
experimentally\cite{Li 90,Reznikov 95,Kumar 96,Kurdak 96,vanderBrom
98} (see also Ref.~\onlinecite{Beenakker 91}). All these works focused on the
suppression of noise by Fermi correlations, which is especially strong
when the channel transparency is close to 1. In order to observe
this suppression, the experimental
studies were invariably performed in a ''quantum point contact''
geometry, i.e., within conductors with a small number of quantum
channels, at temperatures much lower than the energy separation between
those channels. (The case of noise in multimode ballistic Fermi systems
did not attract that much attention.)

However, shot noise at ballistic transport may also be suppressed by
Coulomb interactions of an electron with the space charge of other
ballistic electrons. For nondegenerate, Boltzmann electron gas this
effect was well studied long ago, mostly in the context of vacuum tube
noise -- see, e.g., Ref.~\onlinecite{vanderZiel 54} (See also recent Monte
Carlo studies, Ref.~\onlinecite {Gonzalez 97}). To the best of our
knowledge these 
studies have never been extended to the degenerate, Fermi
systems. Such a study is of particular importance for at least two
reasons.

From the fundamental point of view one should wonder whether the Fermi
correlations may have a strong effect on the properties of wide,
multichannel ballistic conductors. As for Coulomb correlations, it can be
argued that its effect on the noise should be regarded as a much more
accessible manifestation of electron-electron interactions than in more
exotic phenomena, e.g., charge density waves.\cite{Landauer
91} In addition, it is important to understand the difference between
nonequilibrium fluctuations in ballistic and diffusive
Fermi systems. In the latter case, noise originates from scattering
inside the conductor,\cite{Beenakker 92,Nagaev 92,Naveh 97} while in
the former case its origin is the randomness of electron emission from
the source. 

From the point of view of possible applications, the
continuing reduction of channel length $L$ of field-effect transistors
will eventually but inevitably lead to devices with ballistic
transport of electrons. A recent approximate analysis of such
transistors has shown that they may retain high performance all the
way down to $L$ $\sim $ 5 nm.\cite{Pikus 97} Such nanoscale MOSFETs
may be used, in particular, as sense preamplifiers in ultradense
nonvolatile memories.\cite{Brown 98,Likharev 98} For these
applications, the shot noise suppression is of principal importance
because it determines how many single-bit memory cells may be served
by one second-level sense amplifier, a number which may 
strongly affect the final memory density.

\begin{figure}[tb] 
\vspace{0cm}
\centerline{\hspace{1cm} \psfig{figure=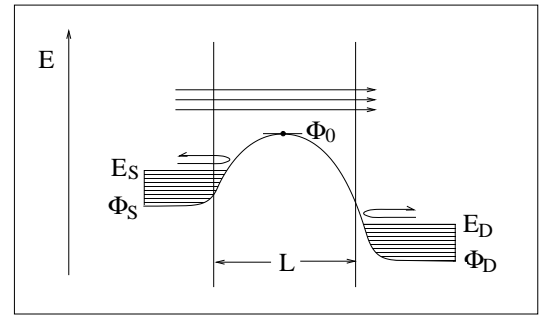,angle=-90,width=70mm}}
\narrowtext
\vspace{0.5cm}
\caption{General scheme of the ballistic Fermi system. \hspace{15cm}}
\label{1profile} 
\end{figure}

The objective of this work is to consider the interplay of the two
mechanisms of shot noise suppression in a broad (multimode) system
with the 
ballistic transfer of electrons from emitter to collector
(Fig.~\ref{1profile}). These 
electrodes are assumed to be in thermal equilibrium; in particular this means
that the collector totally absorbs the incident electrons. (We will revisit
this assumption at the end of the paper.) The potential energy profile
$\Phi (x)$ 
between source and drain, and in particular its maximum $\Phi_0$,  may
be affected by the space charge of ballistic 
electrons. The charge density $en(x)$ is determined, in turn, by the emitted
current.\cite{Pikus 97} 

\vspace{0.0cm}

In order to find the low-frequency shot noise in the device we start with
the basic expression for the spectral density of fluctuations $\delta
i_{S}(E)$ of current $i_{S}(E)$ 
of electrons emitted by the source electrode at some interval $
(E_{t},E_{t}+dE_{t})$ of the total energy $E_{t}$:\cite{Lesovik 89,Buttiker
90,Kogan 96}
\end{multicols}
\widetext
\begin{eqnarray} \label{quantumnoise} 
dS_{is}(\omega ;E_{t}) &\equiv &\overline{\delta i_{S}^{2}(E_{t})}/(\Delta \omega
/2\pi )=2{\frac{e^{2}}{2\pi \hbar }}dE_{t}\left\{ {\rm Tr}({\bf t}^{+}{\bf 
tt{}}^{+}{\bf t})f_{0L}(E_{t})\left[ 1-f_{0L}(E_{t})\right] \right.  
\nonumber  \\
&&\left. +\left[ {\rm Tr}({\bf t}^{+}{\bf t}){-{\rm Tr}(}{\bf t}^{+}
{\bf t{}}{\bf t}^+ {\bf t} )\right] f_{0L}(E_{t})\left[ 1-f_{0R}(E_{t})\right]
\right\}. 
\end{eqnarray}
\vspace{-0.6cm}
\begin{multicols}{2}
\narrowtext
\noindent
Here $f_{0L}(E_{t})$ and $f_{0R}(E_{t})$ are the electron distribution
functions on the left and right hand sides of the interface
between the source electrode and the conductor, respectively. ${\bf t}$ is the
mode transmission matrix across the interface, and the trace is taken over
all the transmission modes 
with energy $E_t$. The assumption of absorptive electrodes implies
that the eigenvalues of ${\bf t}$ are all equal to 1. This, together
with conservation of transversal momentum, allows us to change 
the argument from the total energy $E_{t}$ to the longitudinal energy $
E\equiv E_{t}-E_{\perp }$, and write Eq.~(\ref{quantumnoise}) as
\[	\label{sourcenoise}
dS_{is}(\omega ,E)=2 e i_S (E) [1-f_S(E)],
\]
with 
\[
i_S(E) = {e \over 2 \pi \hbar} N(E) \, dE,
\]
where
\[
N(E)=\int_{0}^{E_{t}}f_{0L}(E+E_{\perp })g(E_{\perp })dE_{\perp }
\]
is the occupancy of transversal modes [$g(E_{\perp })$ is the density of
these modes], and  with
\[
f_{S}(E)=\frac{\int f_{0L}^{2}(E+E_{\perp })g(E_{\perp })dE_{\perp }}{\int
f_{0L}(E+E_{\perp })g(E_{\perp })dE_{\perp }}.
\]
\qquad A similar expression is valid for the intensity $i_{D}(E)$ of
electron emission from the second electrode (''drain'').

In our case of substantial Coulomb interaction between the ballistic
electrons and space charge, we should recalculate fluctuations $\delta
i_{S}(E)$ and $\delta i_{D}(E)$ into those of the total current $I$ . The
general relation between these fluctuations may be written as follows
[we take the direction of 
$i_{D}(E)$ to be from drain to source],\cite{vanderZiel 54}
\[	\label{conversion}
\delta I=\int_{\Phi_S}^\infty \gamma _{S}(E)\delta
i_{S}(E)  - \int_{\Phi_D}^\infty \gamma
_{D}(E)\delta i_{D}(E),
\]
where $\Phi _{S,D}$ are the potentials
deep inside the source and drain, respectively (Fig.~\ref{1profile}) and
factors $\gamma _{S,D}(E)$ describe the space charge screening effect.
In the limit $\omega \tau \ll 1$, where $\tau$ is the ballistic time
of flight,  these factors may be calculated
quasi-statically. In our model, with the source-drain voltage $V
\equiv E_S - E_D$ fixed in time ($E_S$, $E_D$ are the Fermi energies
in the source and drain, respectively),
current $I$ is uniquely determined by $\Phi _{0}.$ This is why we can write: 
\[
\label{gamma}\gamma _{S,D}(E)=\left\{ 
\begin{array}{ll}
{\frac{\partial \Phi _{0}}{\partial i_{S,D}(E)}}{\frac{\partial I}{\partial
\Phi _{0}}} & {\,\,\,\,\,\,\,\,}(E<\Phi _{0}), \\ 
1 \pm {\frac{\partial \Phi _{0}}{\partial i_{S,D}(E)}}{\frac{\partial I}{\partial
\Phi _{0}}} & {\,\,\,\,\,\,\,\,}(E>\Phi _{0}),
\end{array}
\right. 
\]
with the upper (lower) sign used for $\gamma_S$ ($\gamma_D$). In
Eq.~(\ref{gamma}), the 
terms with derivatives describe the effect of the emission current
fluctuation $\delta i_{S}$ on the total current fluctuation $\delta I$ via
that of the local density of electrons, which in turn affects the potential
barrier shape (and $\Phi _{0})$ through the Poisson equation. The unity in
the second line of Eq.~(\ref{gamma}) is due to the direct contribution
of $\delta 
i_{S}(E)$ to $\delta I.$ Now, to calculate $S_{I}(\omega)=\overline{
\delta I^{2}}/(\Delta \omega /2\pi )$ we combine
Eqs.~(\ref{quantumnoise}, \ref{sourcenoise}, \ref{conversion},
\ref{gamma}) and take 
into account the 
statistical independence of sources $\delta i_{S}(E)$ and $\delta i_{D}(E)$
at different energies. As a result, we get 
\begin{eqnarray} \label{SI} 
& &S_{I}(0) =2e\int_{\Phi _{S}}^{\Phi _{0}}dE\, {dI_S(E) \over dE}
\left[ 1-f_{S}(E) 
\right] \left[ {\frac{\partial \Phi _{0}}{\partial i_{S}(E)}}{\frac{\partial
I}{\partial \Phi _{0}}}\right] ^{2}  \nonumber \\
&&+2e\int_{\Phi _{0}}^{\infty }dE\, {dI_S(E) \over dE} \left[
1-f_{S}(E)\right] \left[ 1+ 
{\frac{\partial \Phi _{0}}{\partial i_{S}(E)}}{\frac{\partial I}{\partial
\Phi _{0}}}\right] ^{2}  \nonumber \\
&&+2e\int_{\Phi _{D}}^{\Phi _{0}}dE \, {dI_D(E) \over dE} \left[
1-f_{D}(E)\right] \left[  
{\frac{\partial \Phi _{0}}{\partial i_{D}(E)}}{\frac{\partial I}{\partial
\Phi _{0}}}\right] ^{2}  \nonumber \\
&&+2e\int_{\Phi _{0}}^{\infty }dE\, {dI_D(E) \over dE} \left[
1-f_{D}(E)\right] \left[ 1 -
{\frac{\partial \Phi _{0}}{\partial i_{D}(E)}}{\frac{\partial I}{\partial
\Phi _{0}}}\right] ^{2}
\end{eqnarray}
where $I_S(E)$, $I_D(E)$ are the total source and drain currents at
all energies below $E$, respectively. This equation is 
the main result of our work. It describes the shot noise suppression via
both the Fermi statistics of the emitted electrons (factors
$[1-f_{S,D}(E)]$) and their electrostatic interaction with space
charge (factors with 
derivatives).

\begin{figure}[tb] 
\vspace{0cm}
\centerline{\hspace{0cm} \psfig{figure=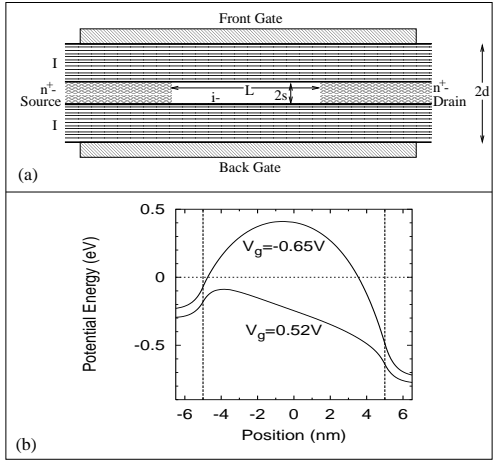,angle=0,width=65mm}}
\narrowtext
\vspace{0.5cm}
\caption{(a) Model of a 2D
ballistic MOSFET and (b) potential energy profiles (relative to the Fermi
level of the source), at high negative and high positive gate voltages.
Vertical lines show position of the conductor-electrode interfaces.
Source-drain voltage is 0.52 V; for other parameters, see the text.}
\label{2model} 
\end{figure}

In order to evaluate the relative importance of these two suppression
factors, we have applied Eq.~(\ref{SI}) to a simple model of a nanoscale,
ballistic, dual-gate MOSFET studied in Ref.~\onlinecite{Pikus 97}
(Fig.~\ref{2model}a). This 
device consists of a thin layer of semiconductor with 2D electron gas
embedded between two gates biased with equal voltages $V_{g}$. The
semiconductor is heavily doped everywhere except in a strip of length $L$
which forms the ballistic channel. Under certain conditions\cite{Pikus 97} the 
potential profile $\Phi (x)$ can be found from the effective 1D Poisson
equation 
\[
\label{Poisson}{\frac{d^{2}\Phi }{dx^{2}}}={\frac{\Phi +eV_{g}}{\lambda ^{2}}
}-{\frac{4\pi e^{2}n}{\kappa _{1}},}
\]
where $\lambda ^{2}=s^{2}/2+(\kappa _{1}/\kappa _{2})s(d-s)$, $2s$ and $2d$
are the semiconductor thickness and the total distance between the gate
electrodes, respectively, and $\kappa _{1}$, $\kappa _{2}$ are the
dielectric constants of the semiconductor and the insulator, respectively.
The simple equation (\ref{Poisson}), together with the assumption of thermal
equilibrium in source and drain, makes possible a semi-analytic scheme for
finding all the quantities of interest. In that scheme the calculation
starts with fixing $\Phi _{0}$ while $I$, $\Phi(x)$, and $L$  are
calculated by simple 
integration, enabling the final calculation of $I$ and $\Phi (x)$ for any $L$
by numerical interpolation. Figure~\ref{2model}(b) shows the
calculated potential 
energy profile for large positive and negative gate voltages $V_{g}$, for a
device with the following parameters: $2s=1.5$ nm, $2d=6.5$ nm, $L=10$ nm, $
\kappa _{1}=12.9$, $\kappa _{2}=3.9$, $T=300K$ and donor density in the
source and drain $N_{D}=3\times 10^{20}{\rm cm}^{-3}$. (With the effective
electron mass of 0.2 $m_{0}$, the latter parameter means that the Fermi
energy in the electrodes is close to 0.3 eV.) \ This parameter set
corresponds to a Si MOSFET with SiO$_{2}$ gate oxide, optimized for dc
transport properties.\cite{Pikus 97}

\begin{figure}[tb] 
\vspace{0cm}
\centerline{\hspace{0cm} \psfig{figure=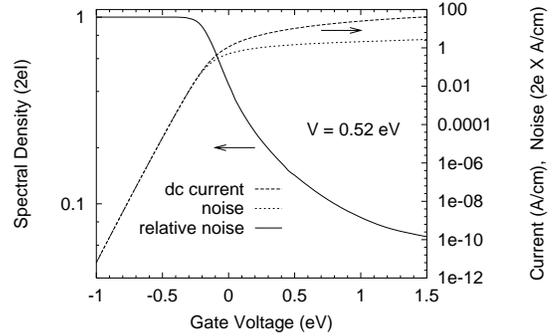,angle=-90,width=70mm}}
\narrowtext
\vspace{0cm}
\caption{DC
source-drain current, and spectral density of its fluctuations (both in
absolute units and normalized to the Schottky value $2eI$) as functions of
gate voltage, at source-drain voltage of 0.52 V.}
\label{3noise} 
\end{figure}

Within the calculation scheme outlined above, the derivatives $
\partial \Phi _{0}/\partial i_{S,D}(E)$ participating in Eq.~(\ref{SI}) may be
found as 
\[
\label{dPhi0dILR}{\frac{\partial \Phi _{0}}{\partial i_{S,D}(E)}}=-\left( {
\frac{\partial L}{\partial i_{S,D}(E)}}\right) _{\Phi _{0}}\left/ \left( {
\frac{\partial L}{\partial \Phi _{0}}}\right) _{i_{S,D}(E)}.\right. 
\]
Figure~\ref{3noise} shows the results of our calculations of the
device noise for 
the parameter set 
specified above. (The results are qualitatively the same for any $L$ between
5 and 15 nm.) At negative gate voltages, when $\Phi _{0}$ is substantially
above $E_S$, current noise is very
close to the 
Schottky value $2eI$; 
however at positive gate voltages the noise may be suppressed by at least
one order of magnitude.

\begin{figure}[tb] 
\vspace{-4cm}
\centerline{\hspace{-2cm} \psfig{figure=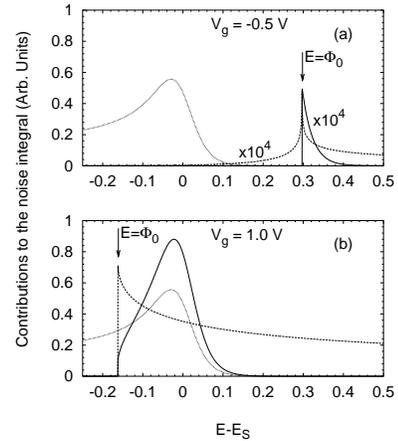,angle=0,width=70mm}}
\narrowtext
\vspace{-1cm}
\caption{The shot
noise suppression factors appearing in Eq.~(\ref{SI}) as functions of energy
at (a) negative and (b) positive gate voltages. Dotted lines show the Fermi
factor, $-\left[ 1-f_S(E)\right] dI_S(E)/dE$, while dashed lines, the
Coulomb factor, $-\left[ d\Phi _{0}/di_{S}(E)\right] dI/d\Phi _{0}$. Solid
line shows the complete integrand [the first term of Eq.~(\ref{SI}) is used
for $E<\Phi _{0}$ and the second term for $E>\Phi _{0}$.]  All the
parameters are the same as in Figs.~\ref{2model} and~\ref{3noise}.}
\label{4components} 
\end{figure}

Figure~\ref{4components} assists the interpretation of this result by
showing the factors 
participating in the first two integrals of Eq.~(\ref{SI}), at two
different gate voltages. Dotted lines represent the Fermi factor which is
concentrated around the Fermi level of the source. The electrostatic term
(dashed lines) peaks at $E=\Phi _{0}$ due to the fact that electrons with
this energy virtually stop near the maximum of the potential profile: $
v(E)\approx 0$. As a result, the corresponding change in charge density, 
$\delta (en)=\delta i_{S}/v(E)$ is very large, and strongly affects
the potential 
distribution. 

At high negative gate voltages [Fig.~\ref{4components}(a)] the Coulomb
interaction 
factor is very small, since
the electron density in the
channel is exponentially small and the space charge effects are negligible.
Also, since $\Phi _{0}$ is well above the Fermi level, the Fermi factor
equals just $dI_S(E) / dE$. This factor gives equal contributions into the dc
current and noise, and so the shot noise has its full, classical Schottky
value. Another way to express the latter fact is to say that at negative
gate voltages the conduction is due to electrons at the upper, Boltzmann
tail of the Fermi distribution, and degeneracy effects are negligible.

At high positive gate voltage when $\Phi _{0}<E_{S}$ 
[Fig.~\ref{4components}(b)], the 
space charge screening factor is rather moderate ($\sim 0.5$) for
energies within the interval $\Phi _{0}<E<E_{S}$ responsible for most
of the current transfer.  The reasons for this weak interaction are
twofold. First, because of electron degeneracy, $I$ depends only
algebraically on $\Phi_0$ (in contrast to the exponential dependence
in the nondegenerate case) so the derivative $\partial I/ \partial \Phi_0$ in
Eq.~(\ref{gamma}) is relatively small. Second, the localized character of the
electrostatic interaction, limited by the exponential cutoff at
distances of the order of $\lambda $ [see Eq.~(\ref{Poisson})] means
that the derivative $\partial \Phi_0 / \partial i_S(E)$ is small. On
the other hand, in this 
open-channel regime, the Fermi suppression of noise is much larger,
because the noise intensity is multiplied by the small term
$[1-f_{S}(E)]$ while the current is not. Physically this means that
while the current is mostly due to all electrons within the interval
$\Phi _{0}<E<E_{S}$, most of these electrons come from far below the
Fermi energy and are ``noiseless''.

Thus, though the device studied in this work is in some sense a Fermi analog
of the classical vacuum tube, its degeneracy and its specific
electrostatics (the exponential 
cutoff of Coulomb interactions of ballistic electrons due to the close
proximity of the gates to the channel) reduce the space charge
effects. The Fermi suppression of the shot noise, however,  is quite strong.
This result is very encouraging from the point of view of applications, but
it may only be considered as preliminary. 

In fact, the simple model of a ballistic MOSFET employed for the
concrete calculations in this work (and used also in
Ref.~\onlinecite{Pikus 97}) 
is somewhat self-contradictory. On one hand, in this model the source,
channel, and drain are all described by the same Poisson
equation~(\ref{Poisson}). This 
implies that not only the channel but also source and drain are 2DEG sheets.
On the other hand, the model assumes perfectly absorptive boundary
conditions at the contact-to-channel interfaces, implying that
the 2DEG in
the contacts remains in thermal equilibrium. This assumption can be easily
justified in geometries where the contacts are bulk (3D), or are made from a
material with a much higher density of states.\cite{Solomon 98} Strictly
speaking, this is not true in our model, if electron scattering in the
contacts obeys the usual hierarchy: elastic scattering events are more
frequent than inelastic ones. In fact, in this case a fraction of ''hot''
(nonequilibrium) ballistic electrons entering the drain will be elastically
scattered back to the channel before they have a chance to thermalize with
the lattice. Thus in order to be realistic for usual devices (including the
Si-based, room-temperature MOSFETs), the model needs to be refined. We
believe, however, that such a modification will not change the results
significantly. 

In summary, we have derived a general equation (\ref{SI}) for shot noise in a
multimode ballistic channel between electrodes in thermal equilibrium, which
describes the noise suppression due to both Fermi correlations of ballistic
electrons and their Coulomb interactions with space charge. Application of
this general result to an approximate model of a ballistic MOSFET shows that
at positive gate voltages the shot noise may be suppressed by more than an
order of magnitude, mostly because of Fermi correlations. Our plans are to
verify this result using more realistic MOSFET models.

We are grateful to P. Solomon for useful discussions and the opportunity to
read his manuscript prior to publication. The work was supported in part by
AME program of ONR/DARPA. 

\end{multicols} 
\end{document}